\documentclass[aps,prl,twocolumn]{revtex4-1}
\usepackage{bm}
\usepackage{amsmath,amssymb,courier,mathtools,blkarray,bigstrut}
\usepackage[]{graphicx,color,epstopdf}
\usepackage{mathrsfs}
\usepackage{textcomp}
\usepackage[normalem]{ulem}
\usepackage{tikz}
\usepackage[T1]{fontenc}
\usepackage{dsfont}

\graphicspath{{figures/}}

\begin{document}


\title{\bf Run stop shock, run shock run: \\Spontaneous and stimulated gait-switching in a unicellular octoflagellate}

\author{Kirsty Y. Wan and Raymond E. Goldstein}
\affiliation{Department of Applied Mathematics and Theoretical
	Physics, Centre for Mathematical Sciences,\\ University of Cambridge, Wilberforce Road, Cambridge CB3 0WA, United Kingdom}
\date{\today}


\begin{abstract}
	In unicellular flagellates, growing evidence suggests control over a complex repertoire of 
	swimming gaits is conferred intracellularly by ultrastructural components,
	resulting in motion that depends on flagella number and configuration. 
	We report the discovery of a novel, tripartite motility in an octoflagellate alga, comprising
	a forward gait (\textit{run}), 
	a fast knee-jerk response with dramatic reversals in beat waveform (\textit{shock}),
	and, remarkably, long quiescent periods (\textit{stop}) within which the flagella quiver.
	In a reaction graph representation, transition probabilities show that gait switching is only weakly reversible.
	Shocks occur spontaneously but are also triggered by direct mechanical contact.
	In this primitive alga, the capability for a millisecond stop-start switch from rest to full speed 
	implicates an early evolution of excitable signal transduction to and from peripheral appendages. 
\end{abstract}

\pacs{Valid PACS appear here}

\maketitle

\smallskip
\noindent

In his \textit{De Incessu Animalium} Aristotle had thus described the walk of a horse \cite{Aristotle}: 
``\textit{the back legs move diagonally in relation to the front legs, for after the right fore leg animals move the left hind leg, 
	and then the left foreleg, and finally the right hind leg}.''
The control mechanism of leg activation was unknown to Aristotle, but is now understood to arise from 
`central pattern generators' \cite{Pearson1993,Holmes2006},
which produce electrophysiological signals (action potentials) that drive limbs in a range of spatiotemporal symmetries.
While microorganisms achieve motility through microscale analogues of limbs called cilia and flagella, 
absent a nervous system the mechanism of control must be quite different.  Nevertheless, 
species of unicellular algae are capable of executing patterns of flagellar beating 
akin to the 
trot and gallop of quadrupeds \cite{Wan2016}. 
In these cases, the extent of intracellular control of appendages
is becoming increasingly evident \cite{Salisbury1988,Wan2016,Tam2015,Klindt2017}.

\begin{figure}[b]
	\centering
	\includegraphics[width=0.8\linewidth]{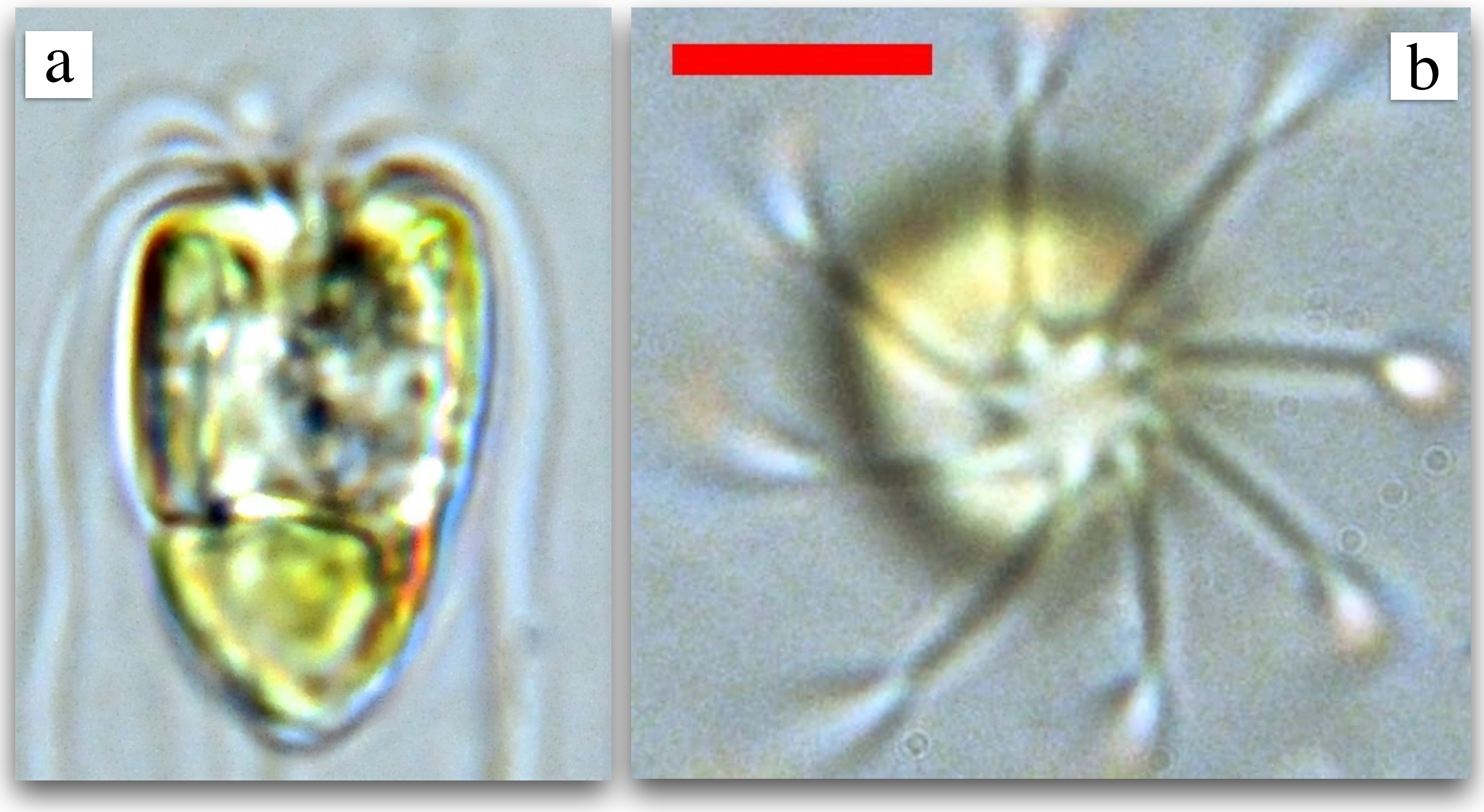}
	\caption{\textit{Pyramimonas octopus}. (a) Side and (b) top views (flagella spiral clockwise). 
		Eyespot visible as conspicuous orange organelle. (Scale bar: $5~\mu$m.)
		\label{fig:1}}
\end{figure}

Here, we detail the discovery of a surprising motility in the 
octoflagellate marine alga \textit{Pyramimonas octopus} (Fig. \ref{fig:1}).
Swimming requires coordination of eight flagella in a pseudo-breaststroke,
in which diametrically opposed pairs beat largely in synchrony.
We find that this forward \textit{run} gait can be interrupted by abrupt episodes involving dramatic changes in flagella beating -- 
hereafter termed \textit{shocks}, which occur spontaneously but can also be induced by external stimuli.
Cells also display a distinctive \textit{stop} gait which can be prolonged,
where cell body movement is stalled but yet the flagella quiver with minute oscillations.
\textit{P. octopus} belongs to a fascinating group of unicellular algae bearing $2^k$ flagella,  
which substantiates a delicate interplay between passive fluid mechanics and active intracellular control in 
the coordination of multiple flagella \cite{Wan2016}. 
Compared to bacteria, the larger size of these algae facilitates visualization, 
allowing us to demonstrate how flagellar beating leads directly to gait-switching and trajectory reorientation,  
and to expose the excitable nature of the eukaryotic flagellum. 

\begin{figure*}[t]
	\centering
	\includegraphics[width=.94\linewidth]{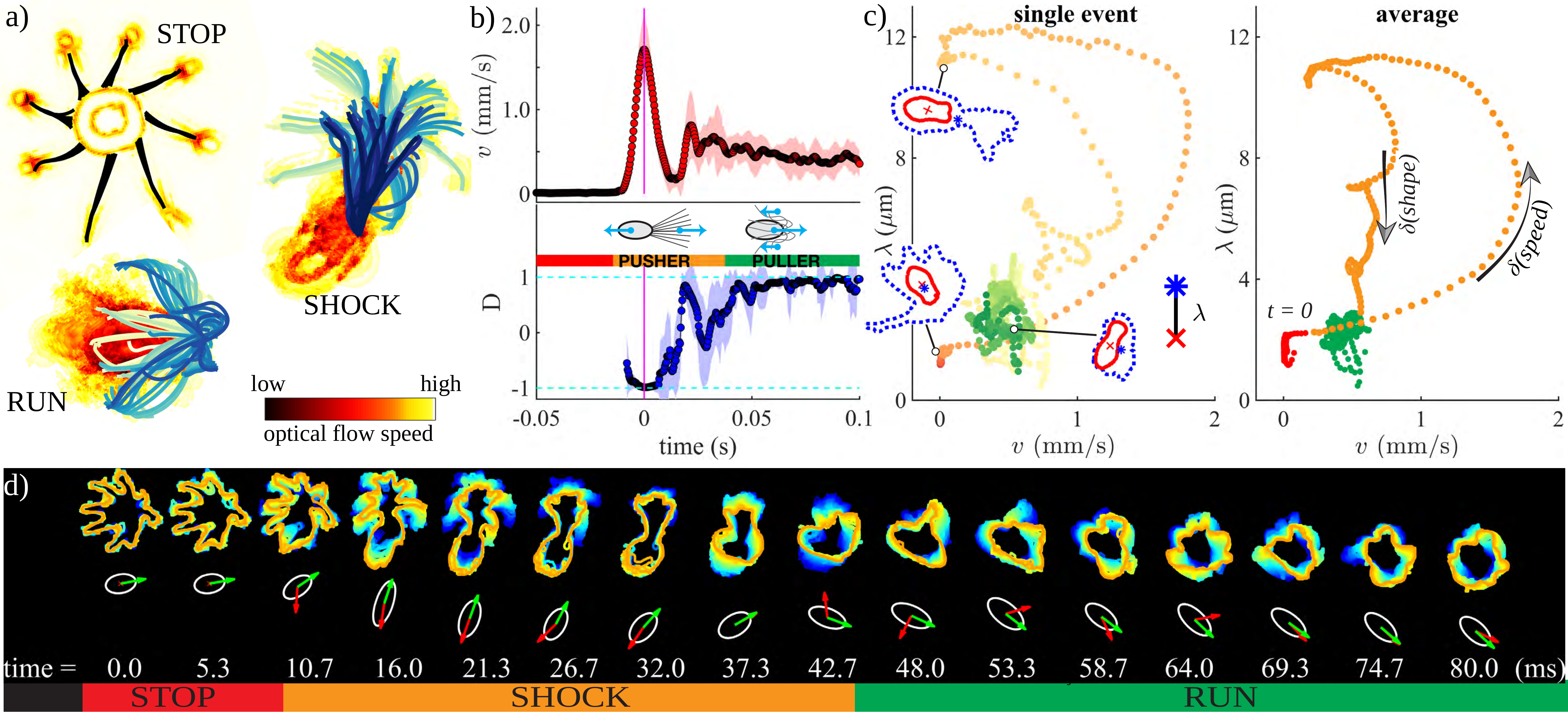}
	\caption{Three gaits of \textit{P. octopus}.
		(a) Cell viewed from the \textsc{top} (\textsc{side}) during stop (run, shock) gaits. 
		Flagella traced from successive frames are overlaid onto optical flow maps following iso-intensity pixels
		(pixel flow rate scales with activity). 
		(b) Transition from stop to run occurs via a shock, with rapid changes in speed $v$ and alignment $D$ (`pusher' to `puller' transition, shaded = 1 standard deviation). 
		(c) Gaits, and the phase trajectories that connect them, are confined to specific regions of  $v$-$\lambda$ phase space.
		(d) Top row: sequence of changes in flagellar beating and cell orientation, plotted here on two timescales (coarse $10$ ms, fine: $5$ ms).
		Bottom row: ellipses -- cell body,  arrows -- cell orientation $\hat{\mathbf{e}}_R$ (red), direction of motion $\hat{\mathbf{v}}$ (green).
		\label{fig:2}}
\end{figure*}

Cell cultures were obtained from the 
Scandanavian Collection of Algae and Protozoa (SCCAP K-$0001$, \textit{P. octopus} Moestrup et Aa. Kristiansen $1987$),
and grown in Guillard's F$/2$ medium under controlled illumination (14:10 day/night diurnal cycle, at $22^\circ$C).
Cells are oblong or rectangular in aspect (Fig. \ref{fig:1}), with
length ($17.05\pm1.74~\mu m$) and width ($9.05\pm1.23~\mu m$).
In their vegetative state cells have $8$ flagella, each of length comparable to the longitudinal dimension of the cell body,
which emerge radially from an apical grove \cite{Moestrup1989}. 
Imaging was conducted under white light illumination on an inverted microscope (Nikon Eclipse TE$2000$-U) and 
high-speed recordings made at up to $3000$ fps (Phantom v$311$, Vision Research). 
Organisms were harvested during exponential growth (at $10^4 - 10^5$ cells/cm$^3$),
and $50-150~\mu$l of suspension were pipetted gently into shallow quasi-$2$D chambers 
(top + bottom: glass, side: Frame-Seal slide chambers -- BIO-RAD) 
for imaging and precision cell and flagella tracking
via custom \textsc{Matlab} algorithms and \textsc{ImageJ} extensions -- see Supplemental Materials (SM).
We ensured cell viability by minimizing environmental stress responses: acclimating cells prior to 
observation, and limiting continuous light exposure to $\lesssim 15$ minutes.

When swimming freely, cells spin about their long axis.
Restricting to individuals traversing the focal plane, we can observe the flagella distinctly.
The \textit{run}, \textit{shock}, and \textit{stop} gaits (Fig. \ref{fig:2}a) are coincident with the three major modes of beating, 
respectively (ciliary, flagellar, and quiescent) \cite{modes}. 
Changes in flagellar activity produce gait transitions.  
However, unlike their bacterial counterparts, eukaryotic flagellar beating is not due to basal rotors but 
rather a coordinated action of dyneins distributed throughout the axoneme \cite{Sartori2016}.
Forward swimming in \textit{P. octopus} arises from ciliary beating (`puller'), 
but during shocks all eight flagella are thrown abruptly in front of the cell where they undulate in 
sperm-like fashion (`pusher'). 
Significant hydrodynamic interactions synchronize the flagella during shocks. 
These `knee-jerk' reactions last only $20-30$ ms, and are related to 
the escape response of \textit{Chlamydomonas} and \textit{Spermatozopsis}.
The latter is triggered by intense photo- \cite{Eckert1972,Witman1993} or mechanical stimuli \cite{Fujiu2009},
but last much longer ($0.2-1.0$ s) and do not occur spontaneously. 
The stop gait has no equivalent in the repertoire of green algae studied so far.

\begin{figure*}[t]
	\centering
	\includegraphics[width=.94\linewidth]{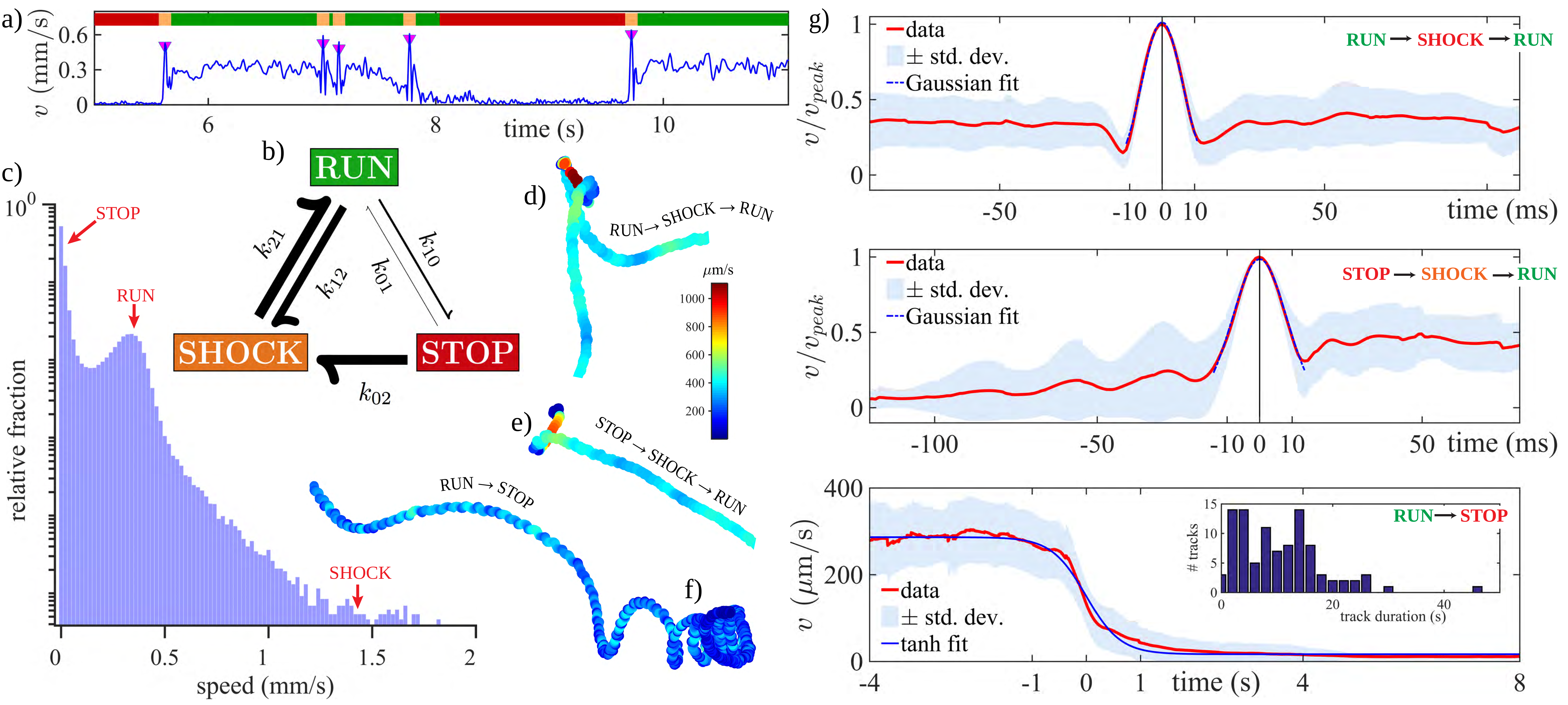}
	\caption{Gait transitions. 
		(a) Instantaneous speed $v(t)$ is partitioned into three states ($0$: stop, $1$: run,. $2$: shock).
		Peaks, at downward triangles, correspond to shocks. 
		(b) Permissible gait bifurcations are indicated by arrows (weighted by transition probability).
		(c) Probability density distribution of speeds indicates proportionality of dwell times in each state. 
		(d)-(f) Trajectories of $3$ characteristic transition sequences. 
		(g) Superimposed and averaged $v$-timeseries exhibit pulse-like maxima when shocks are involved, 
		but much longer decay if converting from runs to stops. Inset: histogram of track durations.
	}
	\label{fig:3}
\end{figure*}
We focus on the stereotypical sequence \textit{stop} $\rightharpoonup$ \textit{shock} $\rightharpoonup$ \textit{run}:
a cell initiates a run from rest via a shock (Fig. \ref{fig:2}b).
Defining the instantaneous alignment $D= \hat{\mathbf{v}}\cdot\hat{\mathbf{e}}_R$ between the 
swimming direction $\hat{\mathbf{v}}$ and the cell body axis 
$\hat{\mathbf{e}}_R$, the puller-like run ($D=1$) may be distinguished from the pusher-like shock ($D=-1$).
Averaged over $10$ cells,
the translational speed rises rapidly from zero to a maximum of $1,712\pm392$ $\mu$m$/$s, but
relaxation to a mean run speed of $428\pm 64$ $\mu$m$/$s takes $\sim0.05$ s.
To separate the flagellar motion from body orientation, we track two dynamically morphing 
boundaries that are delineated by image intensity: 
an inner one for the cell body, and an outer one exterior to the flagella (SM).
The lengthscale $\lambda(t)=\bigg\lVert\sum_{\mathbf{x}\in{\cal B}\setminus {\cal A}} \mathbf{x}/|{\cal B}\setminus {\cal A}| 
- \sum_{\mathbf{x}\in {\cal A}} \mathbf{x}/|{\cal A}| \bigg \rVert$, 
measures the physical separation between the flagella and the cell body proper, where $||\cdot||$ is the Euclidean norm, 
$|\cdot|$ the cardinality of a set, 
and ${\cal A}$, ${\cal B}$ are pixels interior of the inner and outer boundaries respectively.
Naturally, cells at rest exhibit minimal shape fluctuations.
In Fig. \ref{fig:2}c, the three states (realized at instants $t = t_1$, $t_2$, and $t_3$), localize to specific regions of phase-space.
Averaging over multiple events, bifurcations from stops to runs via shocks appear as loops with 
two distinct branches, one involving rapid changes in \textit{speed}, and the second in \textit{shape} (Fig. \ref{fig:2}c).

To estimate the transition probabilities between gaits, we implemented a continuous time Markov model,  
where the instantaneous speed $v$ was discretized to automate a three-state classification from the empirical tracking data (Fig. \ref{fig:3}a).
The state variable $X(t)$ takes the values $\{0 = \text{stop},~1 = \text{run},~2 = \text{shock}\}$. 
The transition rate matrix $ Q = \{q_{ij}\}$, defined by
$q_{ij} = \lim_{\Delta t\to0} {\cal P}(X(\Delta t)=j | X(0)=i)/\Delta t$ for $i\neq j$ (a time-homogeneous Markov process),
and $q_{ii} = -\sum_{j\neq i} q_{ij}$, was estimated to be:
\begin{equation} \nonumber 
Q = ~~
\begin{blockarray}{c c c c}
\begin{block}{l c c c}
& stop & run & shock & \\
\end{block}
\begin{block}{r [l c r]}
stop~~ \bigstrut[t]  & -0.132~~& 0.008 ~~& 0.124 \\ 
run~~ & 0.281 ~~& -1.329 ~~& 1.049\\
shock~~ & 0 ~~& 19.77  ~~& -19.77  \\
\end{block}
\end{blockarray} 
\end{equation}
(for details, and $95\%$ confidence intervals, see SM).
In total, ${\cal O}(10^4)$s of cumulative recordings (individual track durations $0.5-80$ s) were analysed, 
from which $1,377$ distinct pairwise transitions were obtained from $233$ cells. 
Waiting times were estimated from diagonal entries $-1/q_{ii}$:
stop: $7.60\pm 0.75$ s, run: $0.75\pm 0.03$ s, shock: $0.05\pm 0.002$ s (uncertainties are std. errors).
The process admits an embedded Markov chain for discrete jump times, 
with entries $\{k_{ij}, i\neq j\}$ analogous to chemical reaction rates, which represent the probability of transitioning from $i\to j$ 
conditioned on a transition occurring ($\sum_j k_{ij}=1$, $\forall i$).
Here $k_{ii}=0$ (no self-transitions), and 
$k_{01} = 0.0582, k_{02} = 0.9418, k_{10} = 0.2112, k_{12} = 0.7888, k_{20} = 0$ and $k_{21} = 1.0000$ (Fig. \ref{fig:3}b).
Every state is positive recurrent and the process is irreducible.
While run $\rightleftharpoons$ shock bifurcations occur readily, the direct reaction shock $\rightharpoonup$ stop is not possible.  
The network is weakly reversible, not reversible \cite{Anderson2015}, and detailed balance is clearly violated 
(as is the Kolmogorov flux criterion: $k_{01}k_{12 }k_{20} \neq k_{02}k_{21}k_{10} $).
The model predicts an equilibrium distribution $\pi(\text{stop}, \text{run}, \text{shock}) = (0.6666, 0.3126, 0.0208)$.
From a histogram of speeds (for a larger dataset which also includes tracks with no transitions), we estimated the relative dwell times in each 
state:  ($68.6\%,~30.8\%,~0.6\%$), according to cut-offs of $0\sim40$, $40\sim500$, $>500$ $\mu$m$/$s (Fig. \ref{fig:3}c),
which is similar to $\{\pi_i\}$: with discrepancies arising due to subjectivity in choice of cut-off, and prevalence of short-duration tracks.

Gait-switching can greatly affect free-swimming trajectories.
Fig. \ref{fig:3}d-f zooms in on three primary sequences permitted by Fig. \ref{fig:3}b: 
run $\rightharpoonup$ shock $\rightharpoonup$ run, stop $\rightharpoonup$ shock $\rightharpoonup$ run, and
run $\rightharpoonup$ stop.
Typically for photosynthetic unicells, forward swimming is helical with a variable pitch superimposed onto self-rotation. 
Tracks comprise low-curvature portions due to runs, and sharp turns due to rapid conversion 
of flagellar beating and transient reversal during shocks (Fig. \ref{fig:2}d).
Canonical runs decelerate from $\sim400$ $\mu$m$/$s to full-stop, by sequentially deactivating \textit{subsets of flagella} (SM),
the ensuing torque imbalance gradually increases track asymmetry and curvature (Fig. \ref{fig:3}f).
Gait-switching requires two very disparate timescales (Fig. \ref{fig:3}g): 
an ultrafast, millisecond, timescale for bifurcations to and from shocks, but a much slower one for entry into stop states.
The former is reminiscent of neuronal spiking while the latter is akin to decay of leakage currents.
For the first two sequences, the mean is well-fit to a sharply peaked Gaussian ($\sigma=8.6$ ms, $11.6$ ms respectively),
whereas run to stop conversions follow a switch-like profile $A \tanh\left[(x-x_0)/\tau\right]$ with $\tau = 640$ ms. 

\begin{figure}[t]
	\includegraphics[width=\linewidth]{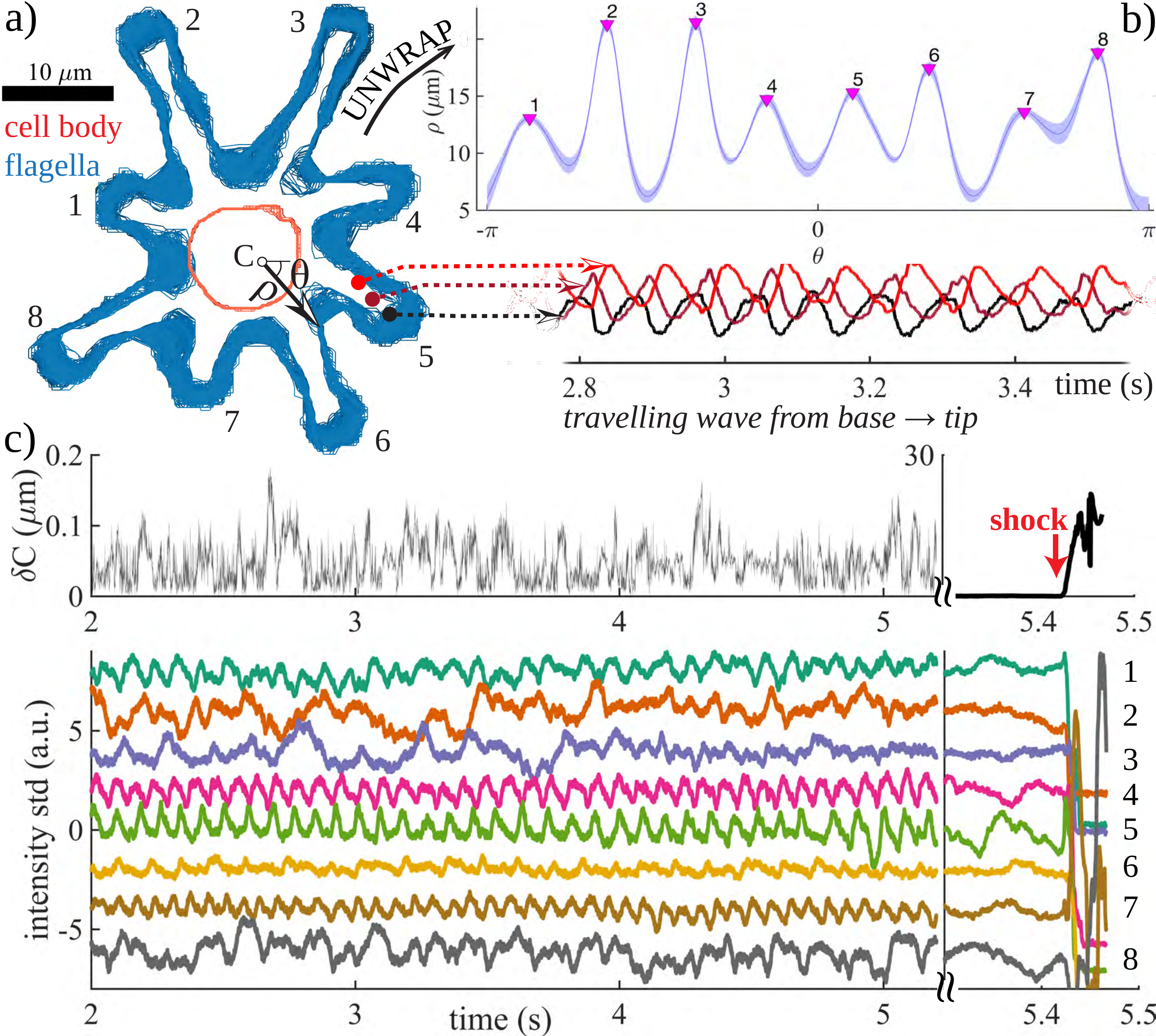}
	\caption{ Arrested, yet still `beating'.
		a) Cell boundaries (with or without flagella) in the stop state are tracked. 
		b) The unwrapped flagellar envelope exhibits $\mu$m-fluctuations.
		Travelling waves, inferred from image intensity changes, propagate outwards from base to tip (inset).
		c) Centroid fluctuations are sub-pixel, random, but flagellar tips display robust oscillations (e.g. flagella 4,5).
		Later, activation of full-amplitude beating occurs simultaneously in all flagella (shock).
	}
	\label{fig:4}
\end{figure}
\begin{figure}[b]
	\centering
	\includegraphics[width=1\linewidth]{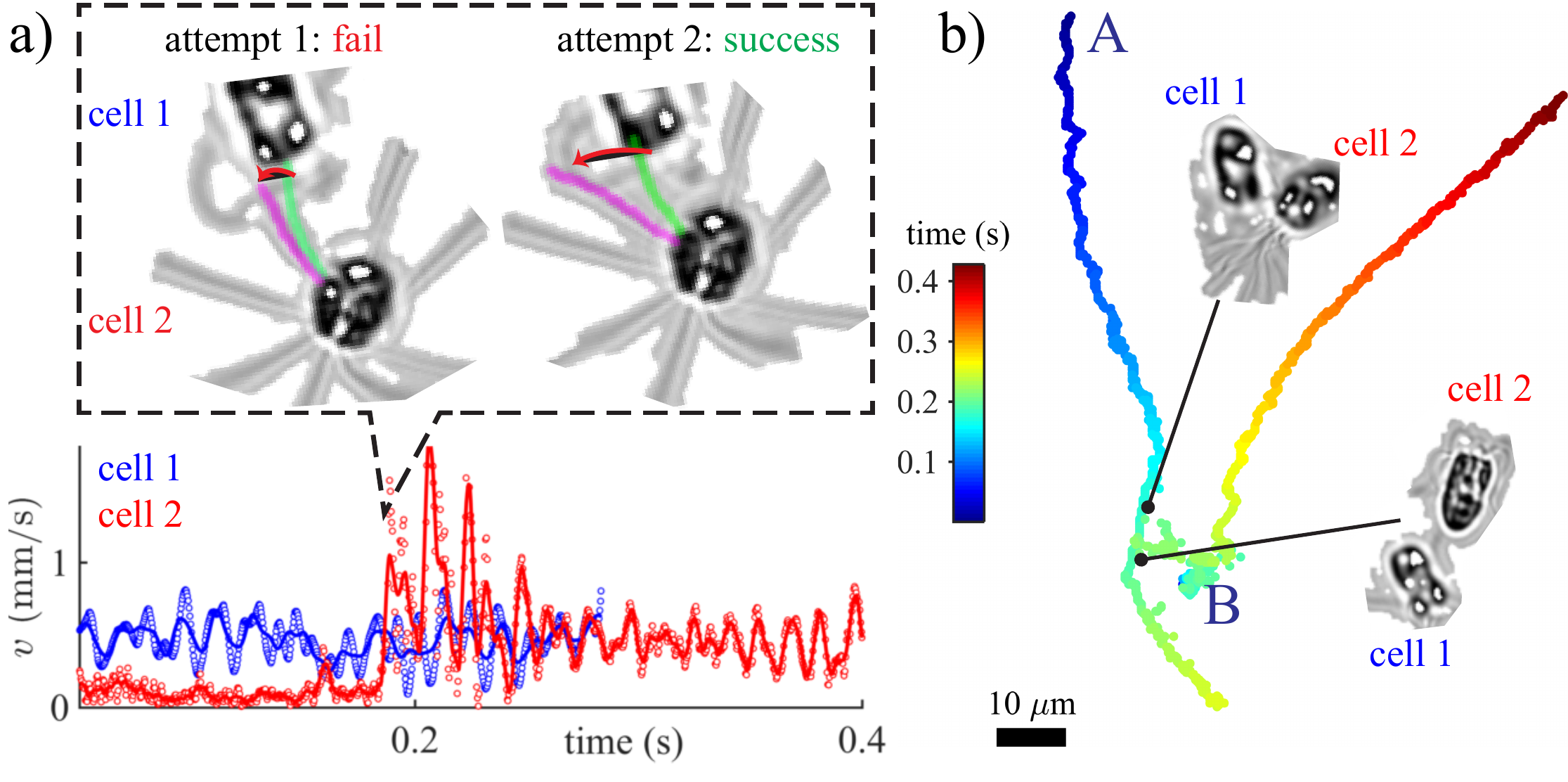}
	\centering
	\caption{Mechanosensitivity: a direct cell-cell collision. 
		a) Contact with only one flagellum is sufficient to trigger a shock, depending on stimulus strength.
		b) Cell $1$ (at A) approaches an initially stationary cell $2$ (at B), induces shock in the latter.}
	\label{fig:5}
\end{figure}

The stopped state can be maintained for up to minutes, before the next restart (Fig. \ref{fig:4}). 
While cell body motion is negligible (sub-pixel variance in centroid displacement: $\sigma_{\delta C} = 0.0253~\mu$m),
significant flagellar activity persists (SM and Movie).
More surprisingly, we deduce using optical analysis that flagellar tip fluctuations can even be oscillatory!
This highly unusual mode may be related to (the much faster) hyperoscillations of reactivated sperm flagella, 
where the noisy dynamics may be signatures of individual dynein oscillations \cite{Kamimura1989}. 
Emergence of global limit-cycle oscillations in the flagella is Hopf-like.

In addition to effecting directional reorientation and sensing \cite{Peruani2007}, the shock gait serves another key physiological function: 
to enable avoidance of obstacles upon direct mechanical contact.
Compared to \textit{Chlamydomonas}, whose flagella display a certain load-response \cite{Fujiu2009,Wan2014b,Klindt2016,Tam2015},
\textit{P. octopus} possess a much heightened mechanosensitivity, where the same downstream pathways leading to \textit{spontaneous} shocks can be activated by touch, 
to produce \textit{stimulated} shocks that are identical in morphology and dynamics to those described earlier (Fig. \ref{fig:2},\ref{fig:3}). 
Fig. \ref{fig:5} shows an example of a moving cell colliding with a cell at rest:
cell 1 contacts cell 2 multiple times but triggers a shock in cell 2 only when the perturbation is large enough. 
For a non-beating flagellum with bending rigidity $EI = 840$ pN$\mu$m$^2$ \cite{Xu2016} we estimate the contact force $F = 3 EI\cdot{\delta}/L^3$
from the measured tip deflection in the two cases: fail: $\sim3.0$ pN, success: $\sim6.6$ pN.
Signal tranduction from the distal point of contact must have occurred within milliseconds.

The unusual motility of \textit{P. octopus} is a significant departure 
from known classical strategies.
Peritrichous enteric bacteria rotate rigid flagellar helices one way or another to cause runs and tumbles,
producing a two-state, paradigmatic strategy for prokaryotic chemotaxis and gradient sensing
based on stochastic switching between directed swimming and random reorientation \cite{Berg1972}. 
The freshwater alga \textit{C. reinhardtii} displays a eukaryotic version of this,
swimming an in-phase breaststroke \cite{Ringo1967,Guasto2010,Wan2014} 
but turning sharply when biflagellar synchrony is lost (`phase drift') \cite{Polin2009}.
Other bacteria species adopt alternative strategies \cite{Taute2015,Son2013,Xie2011},
e.g. the monotrichous \textit{V. cholerae} undergoes a run-reverse-flick motion where flagellar hook elasticity is key.
Contrastingly, the mechanism of enslavement of \textit{P. octopus} swimming to its flagellar dynamics 
is neither due to motor reversal at the base of the flagellum (as in \textit{E. coli}) \textit{nor} to loss of biflagellar synchronization (as in \textit{C. reinhardtii}), 
but rather to a total conversion of beating waveform along the flagellum axoneme proper.

These algae offer rare insight into the bifurcations between different modes of beating in the \textit{same} organelle.
Identifying principal modes of beating (ciliary, flagellar, or quiescent) with (run, shock or stop) states, 
we adopted a natural framework that is liberated from assumptions of specific prototypical gaits (breaststroke, trot, etc).
Patterns of flagellar actuation even during `run' phases are diverse and species-specific \cite{Wan2016},
and environmental stimuli can elicit further changes \cite{Wan2014b, Kaupp2016}.
The \textit{P. octopus} shock, while identifiable with the stimulus-induced (light, mechanical) avoidance reaction of \textit{C. reinhardtii} and \textit{S. similis}, 
is importantly only one component of a tripartite repetoire, is more than an order of magnitude shorter in duration, and occurs spontaneously.
Our three-state classification therefore does not purport to incorporate the totality of gaits
but rather sheds new light on the physiology of gait control. 
By exploring the statistics of gait transitions, we demonstrated that the gait-switching process, 
and not just flagellar beating itself \cite{Battle2016}, operates far from equilibrium 
-- thereby providing a route to enhanced biological sensitivity \cite{Tu2008}.

The discoveries that flagellar activity in \textit{P. octopus} exhibits rapid activation but slow deactivation, and
that apparently quiescent flagella undergo small-amplitude oscillations, have 
great implications for beat emergence and motor coordination in eukaryotic flagella 
\cite{Oriola2017,Bayly2015,Kaupp2016}.
The millisecond shock timescale facilitates rapid removal from predators or obstacles, 
analogously to the escape response of ciliates \cite{Eckert1972}.
More generally, depending on the species, flagella type, and number, the ways of achieving motion, no motion, or change of motion are diverse.
The purple bacterium \textit{B. photometrium} has a sudden light-induced reaction (the Schreckbewegung reaction, or `fright movement') \cite{Engelmann1883},
whereas sensory inputs that inhibit/enhance the firing rate of a `twiddle generator' \cite{Berg1975} can alter the directionality of bacterial flagellar motors.
The cilia of a more advanced phyllum -- ctenophores -- rely on neurons to switch between oscillatory/non-oscillatory states \cite{Tamm2014}.
As important sensory appendages in animals \cite{Fujiu2009,Delling2016}, rapid transduction of signal must likewise be an essential attribute of mammalian cilia.
Thus, in a very primitive unicellular alga, we may have found an evolutionary precedent for the kind of rapid signalling from a distance that, billions of years hence, 
would come to characterize the key physiological functions of mammalian cilia.
\vskip 5pt

\begin{acknowledgments} Financial support is acknowledged from Magdalene College Cambridge through a Junior Research Fellowship (KYW), and Senior Investigator Award 097855MA from the Wellcome Trust (REG).
\end{acknowledgments}

\bibliographystyle{apsrev4-1}
\bibliography{gaitswitch}

\begin{thebibliography}{36}
\bibitem [{mod()}]{modes}%
  \BibitemOpen
  \href@noop {} {\bibinfo  {journal} {We take ``ciliary'' to mean the
  large-amplitude asymmetric beat exhibited by most cilia (e.g. epithelial)
  where the beat cycle partitions into distinct power and recovery stroke
  phases, and ``flagellar'' to mean the symmetric undulatory beat of sperms}\
  }\BibitemShut {NoStop}%
\end{thebibliography}

\end{document}